# Significantly enhanced second order nonlinearity in domain walls of ferroelectrics

Xuewei Deng, Huaijin Ren, Yuanlin Zheng, Kun Liu and Xianfeng Chen*

*Department of Physics, the State Key Laboratory on Fiber Optic Local Area Communication Networks and Advanced Optical Communication Systems, Shanghai Jiao Tong University, Shanghai 200240, China*

*xfchen@sjtu.edu.cn

Significantly enhanced second order nonlinearity resulting from local symmetry breaking in domain walls is reported. By this new giant nonlinearity, novel hexagonally, conically and trigonally distributed second harmonic beams closely related to domain wall patterns are generated without contribution of reciprocal vectors. Experimental results show that such nonlinearities have made important impact on quasi-phase-matching technique and electro-optic Solc filter, which needs careful reconsiderations.

Domain inversions are of great technological importance for their inversed $\chi^{(2)}$ nonlinearities and unchanged refractive indices, which have been widely used for quasi-phase-matching (QPM) techniques[1] and electro-optic device like periodically poled lithium niobate (PPLN) Solc filters[2]. However, the formation mechanism and the local nature of domain walls (DWs) are still far from being clearly understood. Recent researches focusing on DWs reveal unexpected characteristics, such as optical birefringence, strain, local electric fields and electromechanical contrast[3] extending from ~5nm to 50μm in the vicinity of DWs. A widely accepted interpretation is that point defects arising from nonstoichiometry of the crystal organize as defect clusters and impose great influence on local structures. Besides, these defects are believed to be strong scatters to visualize DWs by second-harmonic generation (SHG)[4], which is utilized as SHG microscopy. From the view of point-group structures, the antiparallel DWs will lead to intrinsic symmetry breaking in local structures. Processes of making domain inversions also cause lattice distortions near DWs[5], extending such symmetry breaking. Although asymmetries have been reported in DW regions[6-8], there is a lack of direct studies on physical effects of them. Up to now, researches concerning DW regions are still mainly about domain movements[9] and direct interactions between defects and DWs, like pinning and depinning effects[10].

In this letter, we raise the point of significantly enhanced 2$^{nd}$ order nonlinearity existing in DW regions resulting from the intrinsic symmetry breaking, and distinguish those newly generated nonvanishing elements of $\chi^{(2)}$ susceptibility tensor that use to vanish. It's also demonstrated experimentally that not only different types of novel hexagonally, conically and trigonally distributed second harmonic (SH) beams closely related to DWs are generated without reciprocal vectors, but also effects of DWs importantly impact the well-known QPM techniques and PPLN Solc filters. These significantly enhanced $2^{nd}$ order nonlinearities in DWs from symmetry breaking have, as far as we know, never been reported and systematically considered, while the experimental results reported here impose great challenges on previous studies and strongly recommend that influences of such localized enhanced nonlinearities on QPM processes and electro-optic device like PPLN Solc filters should be restudied more carefully.

The form of $2^{nd}$ order nonlinear susceptibility tensors is constrained by the spatial symmetry properties of the crystals[11]. Under Kleinman symmetry condition, the nonlinear susceptibility tensor can be written as,

$$d_{ij} = \begin{bmatrix} d_{11} & d_{12} & d_{13} & d_{14} & d_{15} & d_{16} \\ d_{21} & d_{22} & d_{23} & d_{24} & d_{25} & d_{26} \\ d_{31} & d_{32} & d_{33} & d_{34} & d_{35} & d_{36} \end{bmatrix}. \quad (1)$$

Increasing spatial symmetry of the crystal imposes more restrictions on the form of the susceptibility tensor. For example, the $LiTaO_3$ and $LiNbO_3$ samples used in our experiment belong to the 3m point group, so only 8 nonvanishing elements exist, 3 of which are independent ($d_{15}=d_{24}=d_{31}=d_{32}$, $d_{16}=d_{21}=-d_{22}$, $d_{33}$). In other words, breaking the crystal's intrinsic symmetry structure will change existing elements and generate new nonvanishing elements. It has also been pointed out by Shen that highly asymmetric molecules with strong charge-transfer bands appear to yield large $|\chi^{(2)}|$ if the crystal structure is also highly asymmetric[12]. Surface SHG in centrosymmetric crystals demonstrates this nonvanishing nonlinear susceptibility due to symmetry breaking of the surface[11]. Analogously, antiparallel DWs break the intrinsic symmetry in local area and extend the scale of asymmetry by lattice distortions. This has been demonstrated quite recently through Raman scattering[13]. It is important to define precisely the meaning of a DW width containing such distortions. According to previously reported result, the appropriate width is limited below 100 nm[7].

A simplest model is a single DW with antiparallel domains adjacent to it. We assume the presence of a special nonlinear susceptibility $\chi_i^{(2)}$ in the thin DW region. As the schematic in Fig. 1(a), enhanced nonlinearity leads to SHG along the wall direction and Cherenkov SHG (CSHG) in the peripheral directions[14, 15]. Conically emitted CSHG in single-domain $LiNbO_3$ has been reported in very early days[16], where the radiating source is isotropic in transverse directions to the FB. In this thin DW model, the geometry is planar and the nonlinear polarization is coherent over the full width of the beam. Therefore, the CSHGs should emerge as well-collimated beams symmetrical to the DW. In the experiment, we fabricate domain inversion with high-quality DWs in a z-cut 1mm $LiTaO_3$ sample (Fig. 1(b) gives the +z surface image). The light source is a Ti: Sapphire oscillator producing about 100fs pulses at wavelength 800nm with averaged power 240mW. 100mm lens is used to focus the FB to 10μm on the input facet. A pair of well-collimated CSHGs symmetrical to DW is observed, as shown in Fig. 1(c). Considering the size of FB spot on the DW, the effective excited wall width is estimated to be about 10nm, in good agreement with earlier papers[3, 7]. The central on-axis SH is blocked when eliminating the influence of the FB and will be shown in the following experiment. We note that a similar result has been reported in PPKTP[15]. However, the author attributed the nonlinearity to dc piezoelectric field caused by strain in domain wall region. We carefully examine their and our experimental results and exclude their explanation with caution. There are three main reasons: firstly, strain can be relaxed by annealing the sample, while there are no obvious differences before and after annealing in our experiment; secondly, $d_{24}$ in Ref. 15 which is used to generated the reference CSHG has indeed been significantly enhanced, because no CSHG is observed in single-domain KTP; and thirdly, instead of the dark line, enhanced SHG in domain wall is observed in our experiment. So we attribute this significantly enhanced $2^{nd}$ order nonlinearity in DWs mostly to intrinsic symmetry breaking.

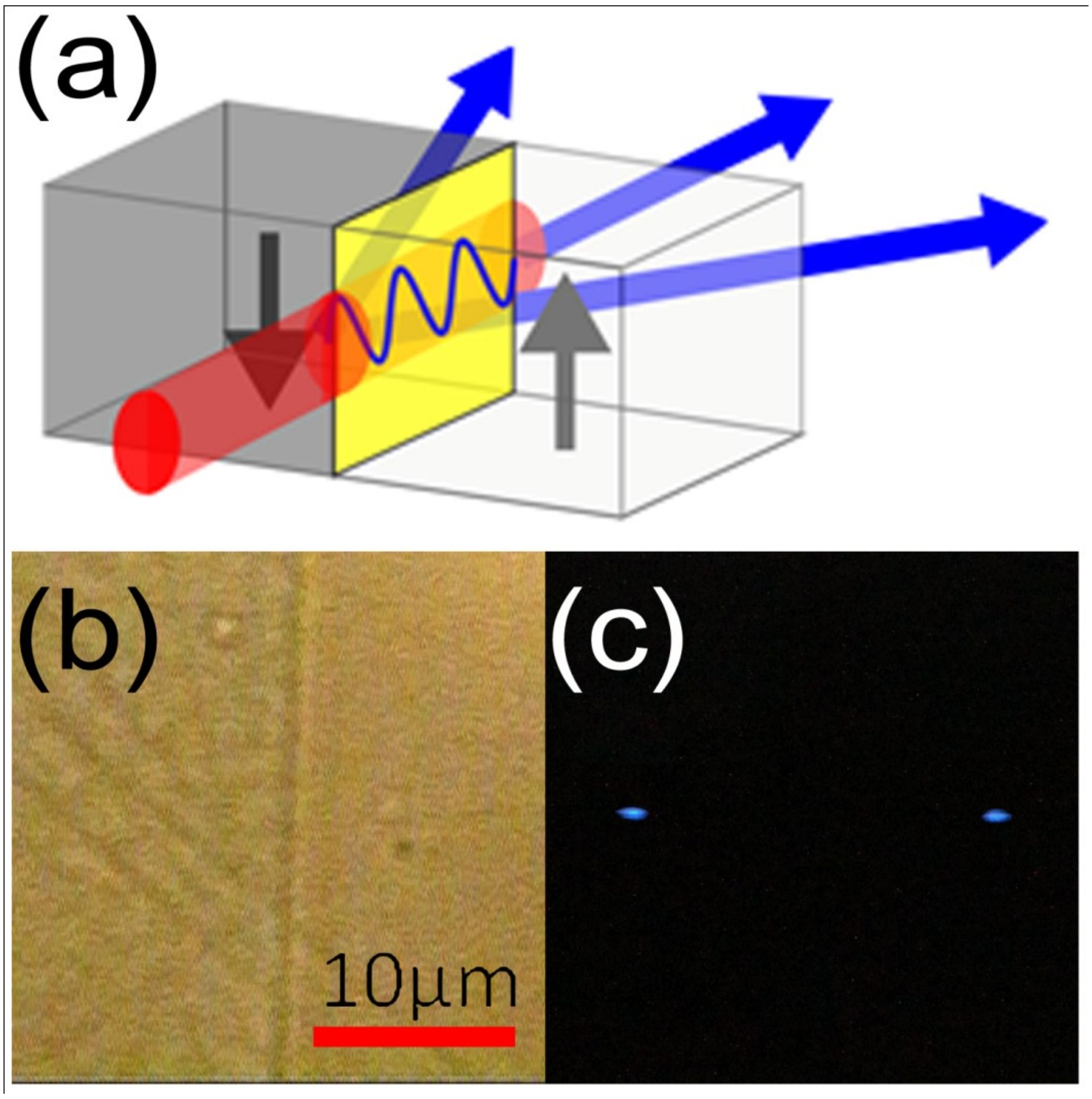

Fig. 1 (a) Schematic of single DW model; (b) DW with high quality on +z surface of $LiTiO_3$ sample; and (c) corresponding CSHG pattern to single DW.

To apply this single wall model to more complex wall patterns, we fabricate two kinds of micro domain inversions on the sample as illustrated in Fig. 2(a) and (b). The first kind possesses lateral DWs parallel to z axis. The second, which does not reach –z surface and possesses tilted lateral DWs, can generate SH beams along DWs in the off-axis directions so that these SH beams can be observed than being blocked on-axis. A third kind area with complex DW pattern is also found as shown in Fig. 2(e). Fig. 2(d) and (f) are corresponding hexagonal and conical CSHG patterns to the DWs shown in Fig. 2(c) and (e). When the FB decreases its diameter to cover two walls or only one wall of the three in Fig. 2(c), the situation is obviously falling back to the single wall model, as shown in Fig. 2(g) and (h). The conical CSHG is an apparent result for the complex DW, where scattering effect cannot be neglected. The aforementioned SHGs along DWs are observed as shown in Fig. 2(i). The trigonal paraxial SH beams well correspond to the trigonal DWs in $LiTaO_3$. The same experiment in $LiNbO_3$ generates hexagonal paraxial SHG due to its hexagonal domain walls. Conical SHG with Cherenkov emitting angle is widely reported in SBN crystals and is interpreted to be phase matched by various reciprocal vectors from randomly distributed antiparallel domain inversions in it[17, 18]. However, according to our experiment, those conical SHGs should be domain-wall CSHGs. Furthermore, novel hexagonal, conical and trigonal SH patterns are generated through simple DWs without any contribution of reciprocal vectors, avoiding two-dimensional (2-D) photonic crystals[19-21]. From our point of view, the peripheral and central SH patterns from those 2-D photonic crystals[19-21] are indeed only relevant to the domain walls that are covered by FB. Reciprocal vectors in these 2-D structures seem less important. The transverse effects in such 2-D structures need cautious revisit.

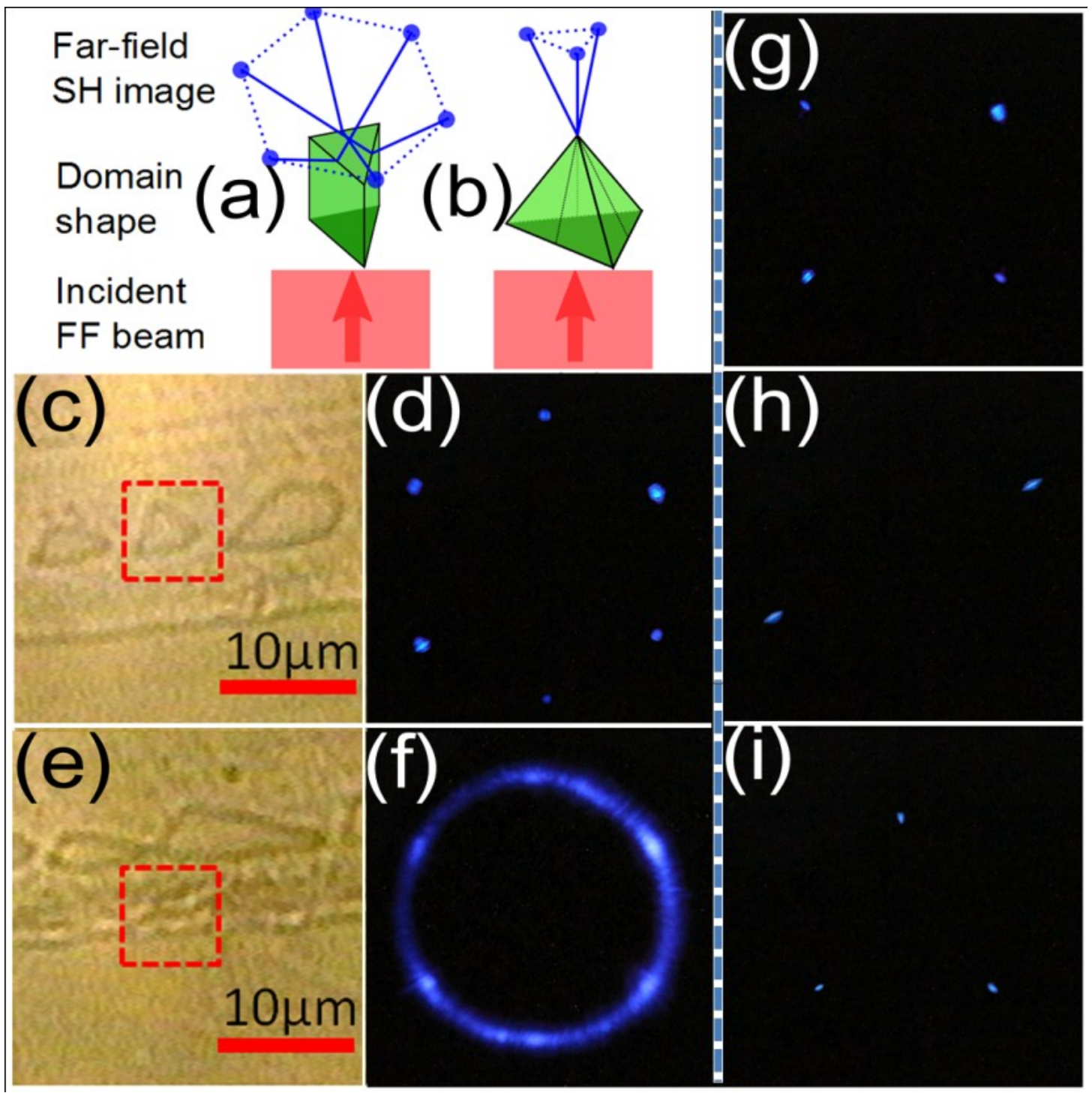

Fig. 2 (a) and (b) Schematic of two kinds of domain inversions and the corresponding CSHG or SHG patterns as described in text; (c) and (e) are DW patterns on +z surface; (d), (f), (g), (h) are CSHG patterns; and (i) is the pattern of SHG along DW direction. See detailed discussion in text.

Polarization analysis reveals new nonvanishing $\chi_i^{(2)}$ elements arising in DWs. Table I gives the results of SH polarizations and involved elements with respect to different FB's directions and polarizations. It shows that compared with bulk $LiTaO_3$, $d_{11}$, $d_{12}$, $d_{13}$ and $d_{23}$ do not vanish in DWs. Enhanced 2$^{nd}$ order nonlinearity provides more radiation sources for CSHG without increasing the source energy[22], so we can simply compare the CSHG efficiencies by repeating the experiment in $LiNbO_3$ as described in Ref. 16. Significant enhancement of $\chi_i^{(2)}$ in DW is found to be about $10^3$. We also measure the CSHG efficiencies of $d_{33}$ and $d_{31}$ separately in DWs of $LiNbO_3$ and get the ratio $(d_{33}/d_{31})_{DW} \approx 1.7$, which is relatively smaller than in the bulk $LiNbO_3$ $(d_{33}/d_{31})_{bulk} \approx 7$. This is another reason why we exclude the proposed explanation in Ref. 15. These results are still valid with slight decline after having annealed the sample at 450 ℃ for 8 h and cooled it down to room temperature for another 8 h. The slight decline may be caused by the defect transition from frustrated state to stable state[3]. To sum up, symmetry breaking in DWs significantly enhances the 2nd order nonlinearities, generates new nonvanishing susceptibility elements and reduces the discrepancies between different elements.

TABLE I. SH polarizations and involved $\chi^{(2)}$ elements with respect to different FB's directions and polarizations.

| FB's direction | FB's polarization | Observed SH's polarization | Involved $\chi^{(2)}$ elements |
|---|---|---|---|
| x | y or z | $E_y$, $E_z$ | $d_{22}$, $\boldsymbol{d_{23}}$, $d_{32}$,$d_{33}$ |
| y | x or z | $E_x$, $E_z$ | $\boldsymbol{d_{11}}$, $\boldsymbol{d_{13}}$, $d_{31}$, $d_{33}$ |
| z | x or y | $E_x$, $E_y$ | $\boldsymbol{d_{11}}$, $\boldsymbol{d_{12}}$, $d_{21}$, $d_{22}$ |

The bold and italic elements are the newly generated ones which use to vanish in bulk $LiTaO_3$.

The most direct consequence of the enhanced 2nd order nonlinearity in DWs, which is the easiest to neglect, is its impact on QPM SHG processes. Although DW is rather narrow compared with the QPM period, its impact on SHG is still not negligible due to the giant $\chi_i^{(2)}$ enhancement of about $10^3$. A 15mm long PPLN sample of 30μm QPM period is used here. Theoretically, 30μm QPM period cannot provide proper reciprocal vectors to phasematch 800nm FB in SHG process. The conversion efficiency as many side bands of sinc function over the FB's spectra (dashed line in Fig. 3 is the calculated envelope of such side bands) will be too low to measure. Experimental result demonstrates the important influence of the giant $\chi_i^{(2)}$ in DWs. As shown in Fig. 3, nearly all the frequency components of FB have been doubled efficiently. This is explicit evidence that such SHG is originated in DWs but not the QPM structure. As a comparison, a 15mm long bulk $LiNbO_3$ is substituted for the PPLN and no efficient SHG is observed. Inset (a) and (b) of Fig. 3 give a direct show of the SH spots. We use three different samples to measure the SH intensities: 10mm long aperiodic PPLN, 15mm long PPLN of 30μm period and 5.8mm long PPLN of 6.5μm period (labeled 1, 2 and 3 in inset (c) respectively). Surprisingly, the SH intensities increase linearly with the DW numbers, as shown in the inset (c) of Fig. 3. The mechanism of such efficient SHG by DWs is complex. An appropriate interpretation is that SHs generated by DWs are randomly phase matched[23, 24] due to the difference between different DWs, leading to a linear increase of SH intensity with the wall numbers. In fact, DW impact on QPM SHG is generally depicted as the decrease of the effective nonlinearity with the decrease of the QPM period[4]. Looking from DW's point of view, this result is obvious. If the QPM period decreases to sub-micrometer and even smaller, the DWs' effect will increase and dominate, where the mechanism of SHG in QPM structures needs to be identified carefully.

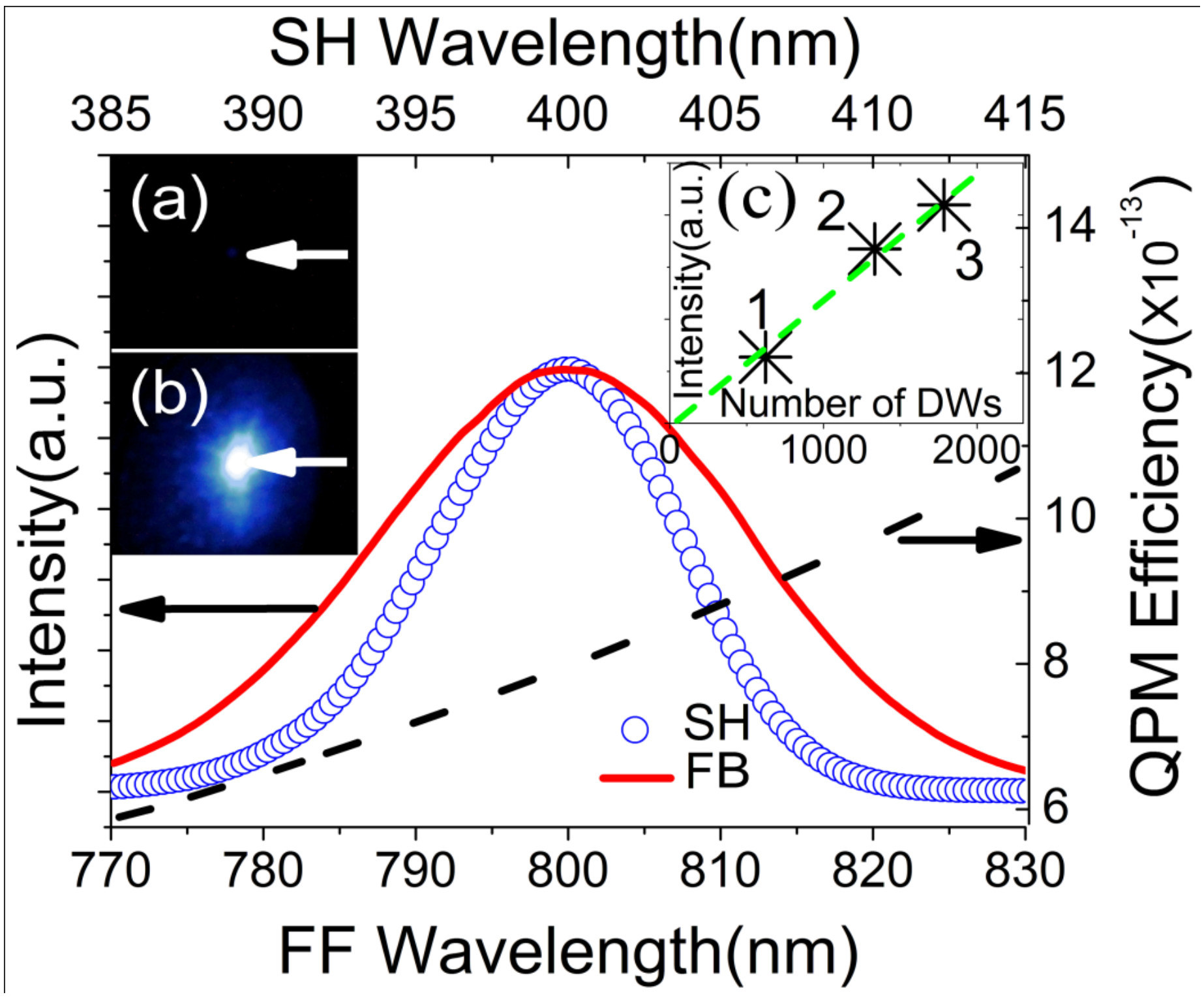

Fig. 3 Spectra of SH and FB. Dashed line is the envelope of calculated side bands of sinc function of QPM efficiency. Inset (a) and (b) are direct comparison of SH spots generated by bulk $LiNbO_3$ and PPLN. Inset (c) shows the linear increase of SH intensities with DW numbers.

Electro-optic susceptibility is one of 2$^{nd}$ order nonlinearities, so DWs' impact is important as well in electro-optic device with DW structures. As in electro-optic PPLN Solc filter[2], only when external electrical field is applied along y axis can the PPLN Solc filter be enabled. Theory without counting DWs' effect predicts that Solc filter can also be enabled without external field when the FB is y-polarized to generate an internal electrical field along y direction through photovoltaic effect[25]. Here, we repeat the experiment in Ref. 2 and find that a z-polarized FB without external electrical field can also enable the Solc filter. The result is shown in Fig. 4. A z-polarized FB will generate internal electrical field in z direction. Since new giant elements arise in the electro-optic susceptibility in DWs, an internal electrical field along z direction can generate electro-optic effect in y direction, which enables the Solc filter. By considering the estimation of nonlinearity enhancement of about $10^3$ in DWs, we calculate the filter effect theoretically[2], which is in good agreement with the experimental result. This Solc filter effect has never been reported before and we attribute this to the significantly enhanced electro-optic susceptibility in DWs.

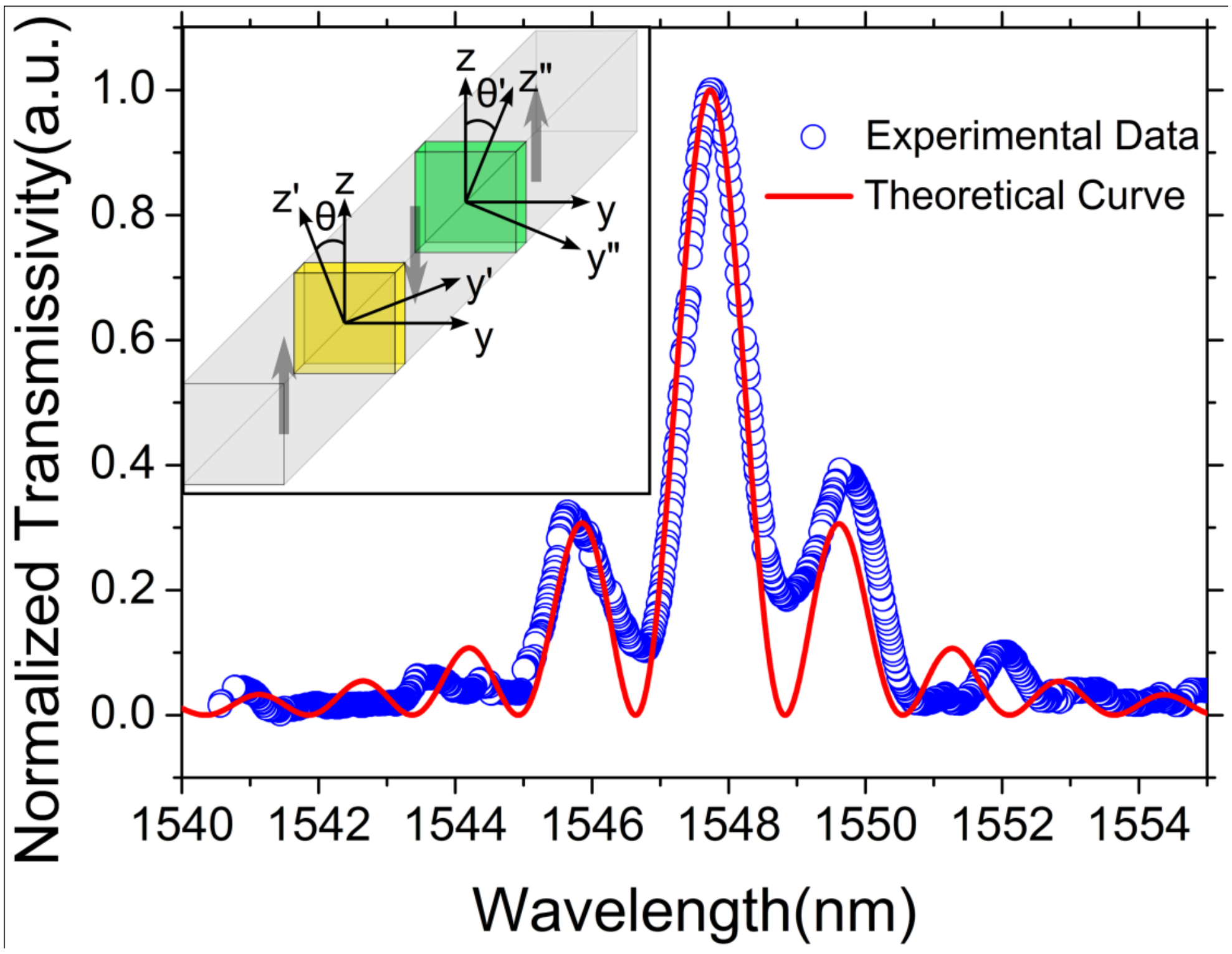

Fig. 4 Experimental result compares with theoretical result of Solc filter by only counting the DWs' contribution. Inset is a schematic of a pair of anti-rocked wave plate formed by DWs when z-polarized FB propagate from positive to negative domain and then from negative to positive domain.

Actually, significantly enhanced 2nd order nonlinearities in DWs due to intrinsic symmetry breaking should exist in all ferroelectrics and a variety of other materials with domain structures. This special characteristic has important influence on novel optical effects, such as harmonic generation, all-optical manipulation, band structure in nonlinear photonic crystals, low-intensity nonlinearity, and so on. Moreover, domain effect will dominate at the scale of sub-micrometer or smaller. So make an intensive study on DWs is vital for research on domain engineering in the future.

In conclusion, we demonstrate significantly enhanced $2^{nd}$ order nonlinearities in DWs from intrinsic symmetry breaking, distinguish those newly generated elements which use to vanish and estimate the enhanced order of magnitude of about $10^3$. Novel hexagonally, conically and trigonally distributed SH beams closely related to DW patterns are generated by the enhanced $\chi_i^{(2)}$ in DWs. Important influences on QPM SHG and electro-optic PPLN Solc filter are discussed and experimentally demonstrated, which have never been studied in previous researches.

This research was supported by the National Natural Science Foundation of China (No. 60508015 and No.10574092), the National Basic Research Program “973” of China (2006CB806000), and the Shanghai Leading Academic Discipline Project (B201).